\documentclass[journal]{IEEEtran}
\usepackage{epsfig,graphicx,subfigure,psfrag,amsmath,cases}
\usepackage{latexsym,amssymb,amsmath,epsfig,subfigure,algorithm,mathtools}
\usepackage{algorithmic}
\usepackage{color}
\usepackage{url}
\usepackage{scrtime}
\usepackage{cite}
\usepackage{hyperref}
\usepackage{subfigure}
\usepackage{bbding}
\usepackage{multicol}
\usepackage{bm}
\usepackage{xcolor}
\usepackage[framemethod=TikZ]{mdframed}
\usetikzlibrary{shadows}
\usepackage{environ}
\usepackage{varwidth}

\newlength{\MyMdframedWidthTweak}%
\NewEnviron{MyMdframed}[1][]{%
    \setlength{\MyMdframedWidthTweak}{\dimexpr%
        +\mdflength{innerleftmargin}
        +\mdflength{innerrightmargin}
        +\mdflength{leftmargin}
        +\mdflength{rightmargin}
        }%
    \savebox0{%
        \begin{varwidth}{\dimexpr\linewidth-\MyMdframedWidthTweak\relax}%
            \BODY
        \end{varwidth}%
    }%
    \begin{mdframed}[
        backgroundcolor=lightgray,
        shadow=true,
        shadowsize=4pt,
        roundcorner=5pt,
        userdefinedwidth=\dimexpr\wd0+\MyMdframedWidthTweak\relax,
        #1]
        \usebox0
    \end{mdframed}
}

\author{Zhiqiang Wei, Weijie Yuan, Shuangyang Li, Jinhong Yuan, Ganesh Bharatula, Ronny Hadani, and Lajos Hanzo \thanks{Z. Wei, W. Yuan, S. Li, and J. Yuan are with the School of Electrical Engineering and Telecommunications, the University of New South Wales, Australia (email: \{zhiqiang.wei, weijie.yuan, shuangyang.li, j.yuan\}@unsw.edu.au);  G. Bharatula is with the Telstra Corp Ltd., Australia (email: Ganesh.Bharatula@team.telstra.com); R. Hadani is with the Cohere Technologies, Inc., and the Mathematics Department, the University of Texas at Austin, USA (email: hadani@math.utexas.edu); L. Hanzo is with the School of Electronics and Computer Science, University of Southampton, Southampton, UK (email: lh@ecs.soton.ac.uk).}}

\title{Orthogonal Time-Frequency Space Modulation: A Promising Next-Generation Waveform}

\newtheorem{T-Prob}{Transformed Problem}

\begin{document}
\maketitle
\begin{abstract}
The sixth-generation (6G) wireless networks are envisioned to provide a global coverage for the intelligent digital society of the near future, ranging from traditional terrestrial to non-terrestrial networks, where reliable communications in high-mobility scenarios at high carrier frequencies would play a vital role.
In such scenarios, the conventional orthogonal frequency division multiplexing (OFDM) modulation, that has been widely used in both the fourth-generation (4G) and the emerging fifth-generation (5G) cellular systems as well as in WiFi networks, is vulnerable to severe Doppler spread.
In this context, this article aims to introduce a recently proposed two-dimensional modulation scheme referred to as orthogonal time-frequency space (OTFS) modulation, which conveniently accommodates the channel dynamics via modulating information in the delay-Doppler domain.
This article provides an easy-reading overview of OTFS, highlighting its underlying motivation and specific features.
The critical challenges of OTFS and our preliminary results are presented.
We also discuss a range of promising research opportunities and potential applications of OTFS in 6G wireless networks.
\end{abstract}

\section{Introduction}

The sixth-generation (6G) wireless networks are expected to support ubiquitous connectivity to a wide range of mobile terminals, spanning from autonomous cars to unmanned aerial vehicles (UAV), low-earth-orbit (LEO) satellites, and high speed-trains, etc.
One of the critical challenges for these services is to provide reliable communications in high-mobility environments.
Additionally, the spectrum congestion under 6 GHz creates a fundamental bottleneck for capacity improvement and sustainable system evolution.
The ultra-high data rate requirements of panoramic and holographic video streaming push mobile providers to utilize higher frequency bands, such as the millimeter wave (mmWave) bands, where a huge chunk of unused spectrum is available.
In general, wireless communications in high-mobility scenarios at high carrier frequencies are extremely challenging due to the hostile channel variations. 
Relying on adaptive coherent/non-coherent detection \cite{XuChaoCoherentNocoherent} is beneficial for attaining a certain degree of robustness against channel variations.
Nevertheless, recently an increasing amount of research attention has been dedicated to designing new modulation waveform and schemes for high-mobility communications of next-generation wireless networks.

High-mobility communications operating at high carrier frequencies suffer from severe Doppler spreads, mainly caused by the relative motion between the transmitter, receiver, and scatterers.
%
%
Conventional orthogonal frequency-division multiplexing (OFDM) modulation, which has been widely used in both the fourth-generation (4G) and the emerging fifth-generation (5G) cellular systems as well as in WiFi networks, suffers in high-mobility scenarios.
OFDM waveform is impaired by severe inter-carrier interference (ICI), which is aggravated by the fact that the highest and lowest subcarriers exhibit rather different normalized Doppler.
Hence, synchronization is also a challenge.
Recently, a new two-dimensional (2D) modulation scheme, namely orthogonal time-frequency space (OTFS)\cite{Hadani2018otfs}, has been proposed as a promising candidate for high-mobility communications.

OTFS modulates information in the delay-Doppler (DD) domain rather than in the time-frequency (TF) domain of classic OFDM modulation, providing a strong delay- and Doppler-resilience, whilst enjoying the potential of \textit{full diversity} \cite{Hadani2018otfs}, which is the key for supporting reliable communications.
Additionally, OTFS modulation can transform a time-variant channel into a 2D quasi-time-invariant channel in the DD domain, where its attractive properties can be exploited.
Given that most of the existing wireless system designs have been conceived for low-mobility and low-carrier scenarios, OTFS introduces new critical challenges  in transceiver architecture and algorithmic designs.
To unleash the full potentials of OTFS, challenging fundamental research problems have to be addressed, including channel estimation, detection, as well as multi-antenna and multi-user designs.

{In contrast to existing tutorial papers on OTFS \cite{Hadani2018otfs,ramachandran2020otfs}, this article portrays OTFS modulation conceived for communications over high-mobility environments by providing an easy-reading overview of its fundamental concepts, highlighting the challenges and potential solutions as well as exploring new promising areas for future research.}
The rest of this article is organized as follows.
The next section introduces the fundamentals of OTFS.
The potential applications and research opportunities of OTFS are presented in Section III.
Then, before concluding, we demonstrate the most important design challenges of OTFS and their potential solutions.

\section{Fundamentals of OTFS}
A fundamental question to answer for motivating the research community and industry to investigate OTFS is why we shall perform modulation in the DD domain.
Hence, commencing from the channel characteristics of high-mobility communications, we present the basic concepts and properties of the DD domain channels, the DD domain multiplexing, the OTFS transceiver architecture and signal waveform, as well as the OTFS system design principle.

\subsection{From Time-Invariant to Time-Variant Channels}
\begin{figure}[t]
	\centering
	\includegraphics[width=3.3in]{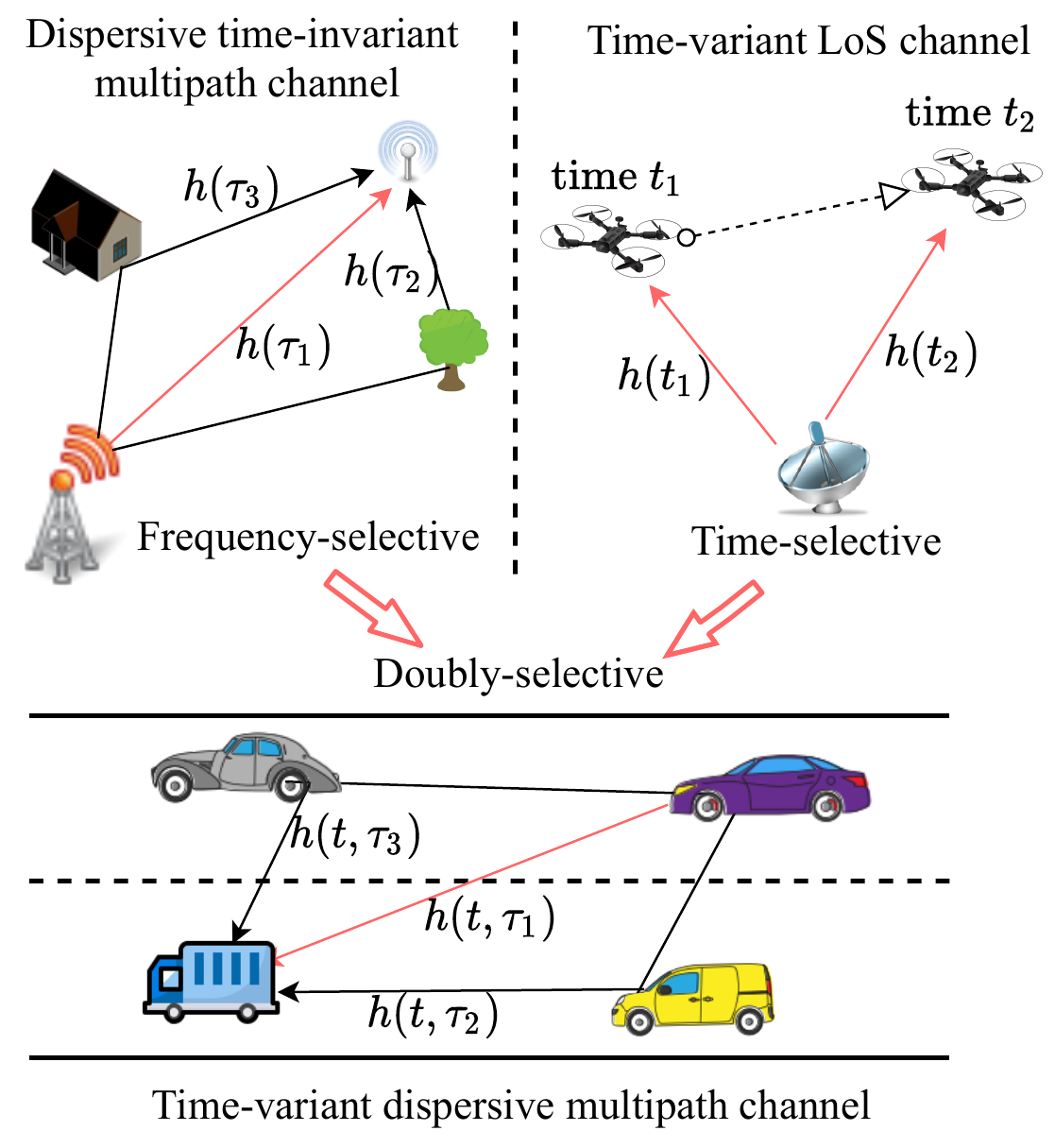}
	\caption{An illustration of frequency-selective, time-selective, and doubly-selective channel models.}
	\label{DoublySelectiveChannels}
\end{figure}

Wireless channels can be modeled by a linear time-invariant (LTI) system, provided that the channel impulse response (CIR) is \textit{time-invariant} or has a long coherence time, cf. Fig. \ref{DoublySelectiveChannels}.
In the presence of multiple scatterers, the dispersive LTI channel's output is a temporal-smeared version of the transmitted signal, but again, the CIR is {time-invariant}.
In this case, a one-dimensional (1D) CIR in the delay domain $h\left(\tau\right)$ is sufficient for characterizing the \textit{time-dispersive} channel.
The Fourier transform (FT) of this CIR is a \textit{frequency-selective} channel transfer function (CTF).
With increasing the CIR delay spread, the selectivity becomes more severe since the separation of frequency-domain (FD) fades is increasingly proportional to the CIR-length.

However, the assumption of having LTI CIRs may no longer hold in the face of increased user mobility and carrier frequency.
Therefore, the linear \textit{time-variant} (LTV) channel model \cite{hlawatsch2011wireless} has attracted considerable research attention in high-mobility scenarios.
LTV channels give rise to frequency shifts due to the Doppler effect, yielding a spectral-smeared version of the transmitted signal, i.e., \textit{frequency-dispersive}.
Frequency-dispersive channels are \textit{time-selective} and the separation of the channel's time-domain (TD) fades is increasingly proportional to the Doppler spread.
In practice, LTV channels of high-mobility scenarios are often \textit{doubly-dispersive} due to the joint presence of multipath propagation and Doppler effects, cf. Fig. \ref{DoublySelectiveChannels}.
The transmitted signals suffer from dispersion both in the TD and FD.
In such scenarios, each tap of the CIR function is time-dependent, fluctuating according to the rate of $\frac{\lambda}{2v}$ between consecutive TD fades, cf. Fig. \ref{TDDomainChannel}, where $\lambda$ denotes the wavelength and $v$ is the relative speed between the transmitter and receiver.
Hence, this results in a 2D CIR function $h\left(t, \tau\right)$ in the time-delay domain.
In contrast to the traditional way of treating TD and FD dispersion as undesired channel impairments, we can beneficially exploit the additional degrees of freedom (DoF) of doubly-dispersive channels for achieving reliable diversity-aided communications in high-mobility scenarios.

\subsection{LTV Channels in TF and DD Domains}
\begin{figure*}[!t]
	\centering
	\includegraphics[width=6.5in]{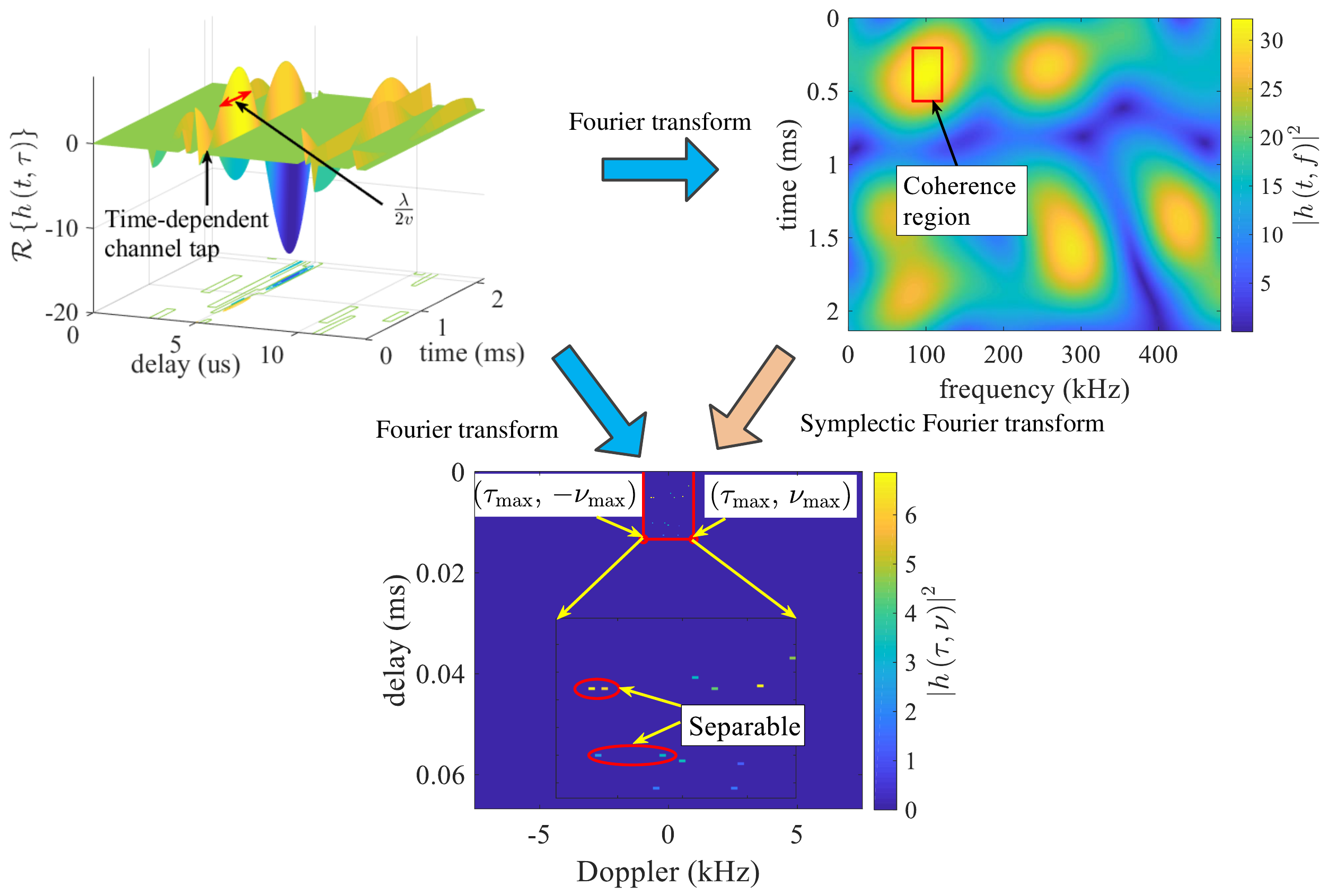}
	\caption{LTV channels in the time-delay, TF, and DD domains.}%
	\label{TDDomainChannel}%
\end{figure*}
	
Apart from the time-delay domain, LTV channels can be equivalently described in either the TF or DD domain, cf. Fig. \ref{TDDomainChannel}.
To emphasize the TF selectivity, the TF domain channel, $H\left(t, f\right)$, can be obtained by the FT of $h\left(t, \tau\right)$ with respect to (w.r.t.)  delay $\tau$.
Note that $H\left(t, f\right)$ can be interpreted as the complex CTF coefficient at time instant $t$ and frequency $f$.
Due to the limited coherence time and coherence bandwidth (coherence region in Fig. \ref{TDDomainChannel}) of LTV channels, channel acquisition in the TF domain would be challenging and would impose a significant signaling overhead.
For instance, for an OFDM system having a carrier frequency of $f_c = 3.5$ GHz and a subcarrier spacing of $\Delta f = 15$ kHz supporting a relative velocity of $v = 300$ km/h, the maximum Doppler shift is $\nu_{\mathrm{max}} = 972.22$ Hz and the OFDM symbol duration including a 20\% cyclic prefix (CP) is $80 \;\mu$s.
Assuming that the channel's coherence time is $\frac{1}{4\nu_{\mathrm{max}}} = 257.14 \;\mu$s, one channel coherence interval can only accommodate at most 3 OFDM symbols.


Applying the FT to $h\left(t, \tau\right)$ w.r.t. $t$ yields the DD domain channel (spreading function), $h\left(\tau,\nu\right)$.
The DD domain channel $h\left(\tau,\nu\right)$ characterizes the intensity of scatterers having a propagation delay of $\tau$ and Doppler frequency shift of $\nu$, which directly captures the underlying physics of radio propagation in high-mobility environments.
More importantly, the LTV channel in the DD domain exhibits beneficial features of separability, stability, compactness, and possibly sparsity, as illustrated in Fig. \ref{TDDomainChannel} and detailed below, which can be exploited to facilitate efficient channel estimation and data detection.
\begin{itemize}
	\item \textbf{Separability}: Additionally introducing the Doppler domain of wireless channels allows us to separate the propagation paths experiencing an identical delay and thus fully discloses the available channel DoF.
	%
	
	\item \textbf{Stability}: Only the drastic change of propagation path lengths and moving speeds may cause channel variations in the DD domain.
	Consequently, DD domain channels fluctuate much slower than time-delay domain or TF domain channels.
	
	\item \textbf{Compactness}: It is worth noting that in typical wireless channels, we have $4\tau_{\mathrm{max}}\nu_{\mathrm{max}} \le 1$ \cite{hlawatsch2011wireless}, where $\tau_{\mathrm{max}}$ indicates the maximum delay.
	Hence, $h\left(\tau,\nu\right)$ has a compact DD domain support within the intervals $\left[0,\tau_{\mathrm{max}}\right]$ and  $\left[-\nu_{\mathrm{max}},\nu_{\mathrm{max}}\right]$ along the delay and Doppler dimensions, respectively.
	
	\item \textbf{Potential sparsity}: When an open-space rural propagation environment having few moving scatterers is considered, the DD domain channel exhibits a sparse response\cite{ShenOTFSCS}.
	%
\end{itemize}

Note that the above discussions are mainly related to the deterministic description of LTV channels.
More details on the stochastic characterization of LTV channels can be found in \cite{hlawatsch2011wireless,SteeleMobileChannel}.
Parameterizing channel with the aid of delay and Doppler is not new, which has been widely adopted in the areas of radar and sonar.
The main benefit of the OTFS waveform is the DD domain multiplexing, which will be detailed in the next section.


\subsection{From TF to DD Domain Multiplexing}

\begin{figure*}[!th]
	\centering
	\includegraphics[width=6.8in]{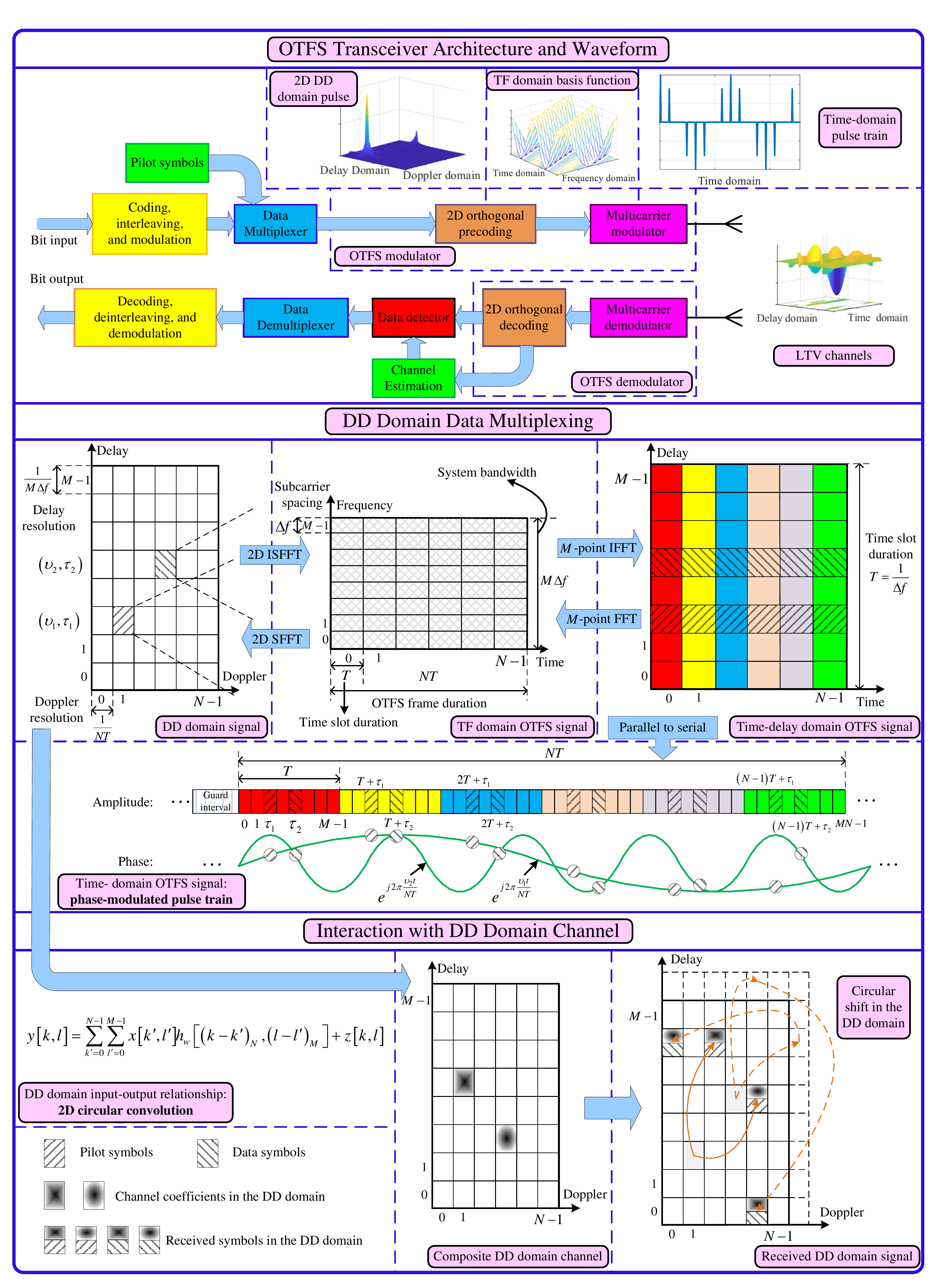}
	\caption{OTFS transceiver architecture, the concept of DD domain data multiplexing and their coupling with DD domain channels.}%
	\label{DDDomainMultiplexing}%
\end{figure*}

The classic modulation techniques typically multiplex data in the TD or FD, which are briefly introduced as follows.
\begin{itemize}
	\item \textbf{Time-division multiplexing (TDM)}: TDM carries information (QAM) symbols in unique, user-specific time-slots, within the same frequency band.
	\item \textbf{Frequency-division multiplexing (FDM)}: FDM multiplexes streams or users in dedicated FD slots occupied at the same time.
	\item \textbf{Code-division multiple access (CDMA)}: The information symbols (of different users) are carried over unique (user-specific) single-carrier TD, multi-carrier FD or multi-carrier time-frequency-domain spreading sequences.
	\item \textbf{OFDM}: OFDM transmits its information symbols by overlapping sinc-shaped orthogonal subcarriers in parallel.
	\item \textbf{Index modulation (IM)}: IM relies on the generalized ON/OFF keying principle to map the information bits to the indices of spatial-, time- and frequency-domain resources.
\end{itemize}
However, these traditional modulation techniques may not work well in the face of severe Doppler spread.
For instance, the popular OFDM modulation efficiently transforms a frequency-selective fading channel to multiple parallel frequency-flat subchannels, which allows a low-complexity single-tap FD equalization for transmission over LTI channels.
However, the orthogonality of OFDM waveform erodes in high-mobility scenarios when conventional wireless transceivers are adopted.


{In contrast to the existing 1D TD or FD modulation techniques, OTFS is a 2D modulation technique, where the data symbols are multiplexed in the DD domain and each symbol is spread right across the entire TF domain.
This property is desirable for attaining the maximum achievable diversity for transmission over doubly dispersive channels, provided that each TD and FD sample experiences independent fading.
The maximum attainable diversity order is determined by the number of independently fading resolvable paths in the DD domain.
To fully exploit the maximum attainable diversity order, advanced detection techniques are required that can efficiently combine the information conveyed by the different propagation paths to the OTFS receiver.
Although the asymptotic diversity gain of the uncoded OTFS system can be as low as one in some special cases as revealed in \cite{SurabhiDiversityAnalysisBER}, the probability of occurrence of these cases tends to zero for a moderate DD domain grid size \cite{RavitejaEffectiveDiversity}.
For example, it becomes as low as $10^{(-150)}$ for a DD domain grid of $M = N =16$ \cite{RavitejaEffectiveDiversity}.
Furthermore, channel coding is capable of avoiding the occurrence of these cases and thus OTFS can achieve the full diversity order almost surely.}

{As a benefit of DD domain multiplexing, the information symbols are carried by localized pulses \cite{Hadani2018otfs,Mohammed2020derivation} of the DD domain.
By increasing the time duration and bandwidth of the transmission, the information-bearing pulses can be further localized, cf. Fig. \ref{DDDomainMultiplexing}, demonstrating the orthogonality of the OTFS waveforms associated with different DD grid points.
This potentially perfect pulse localization capability experienced in the DD domain stands in sharp contrast to the properties of its dual pair, namely the TF domain, where the Heisenberg uncertainty principle prevents corresponding pulse localization.
The above-mentioned pulse-localization allows OTFS to enjoy the aforementioned DD domain channel properties.

Moreover, adopting 2D localized pulses for conveying the information in the DD domain is beneficial for exploiting the underlying physical propagation properties. 
We note that conventional wireless communication designs treat channel fading as an inevitable deleterious effect, aiming for combating or exploiting it while ignoring its basic underlying causes.
In more detail, the channel impairments of propagation delay and Doppler frequency shifts are modeled as a pair of operations imposed by wireless channels on the transmitted waveform. 
Furthermore, fading is viewed as a phenomenon that manifests itself at the channel's output imposed by the combined destructive effect of this pair of fundamental channel-induced phenomena.
In contrast, OTFS generates a complete orthogonal family of waveforms, which is closed under arbitrary combinations of the time delay and frequency shift.
In other words, upon transmitting a signal in this family, the received signal will remain within the family under arbitrary channel impairments. 
The mathematical structure underlying this unique characteristic of the OTFS waveform family is the quasi-periodicity property of the DD domain signal representation, as discussed in \cite{Hadani2018otfs}. 
This property gives rise to a 2D (quasi-) circular localized inter-symbol interference (ISI) pattern in the DD domain, representing the channel impairments, cf. Fig. \ref{DDDomainMultiplexing}. 
Since the time- and frequency-shifts are separated in the DD domain, the destructive effects of fading are substantially mitigated.}

\subsection{OTFS Transceiver Architecture and Waveform}


The OTFS transceiver is shown in Fig. \ref{DDDomainMultiplexing}, where the modulated (pilot) symbols are firstly mapped to the DD domain.
Then, an orthogonal 2D precoding, such as the inverse symplectic finite Fourier transform (ISFFT) and Walsh-Hadamard transform\cite{ZemenOrthogonalPrecoding}, transplants the DD domain signal into the TF domain.
Then, a multi-carrier modulator, such as OFDM or filterbank multicarrier (FBMC) modulator, is employed in each time slot for further transforming the TF domain signal to the TD before being transmitted over the channel. 
At the receiver side, a cascaded combination of multi-carrier demodulation and 2D orthogonal decoding transforms the received signal back into the DD domain and retrieves the transmitted symbols in the DD domain using an appropriate channel estimator \cite{RavitejaOTFSCE} and equalizer\cite{RavitejaOTFS}. 
We can observe that the OTFS transceiver can be implemented based on the conventional OFDM architecture by adding some pre-processing and post-processing blocks, thus making OTFS attractive from the perspective of implementation.
{Nevertheless, as a block modulation scheme, OTFS systems suffer from a higher latency than classic OFDM systems.}

{As illustrated in Fig. \ref{DDDomainMultiplexing}, a single pulse representing a symbol at $\left(\tau,\nu\right)$ in the DD domain will be spread across the whole TF domain.
The resultant TD waveform is the fluctuating pulse train seen in Fig. \ref{DDDomainMultiplexing}, where the fluctuation rate is determined by the Doppler frequency $\nu$, while the pulse location within each time slot is determined by the delay $\tau$.
Consequently, the TD waveform of OTFS behaves locally like TDMA (localized pulses in the TD), globally like OFDM (localized pulses in the Doppler domain) and spreading like CDMA (2D spreading in the DD domain), thus inheriting their beneficial properties.
For example, OTFS exhibits resilience to narrow-band interference and it is eminently suitable for multi-user scenarios.
%
%
%
Additionally, OTFS transforms the violently fluctuating TF domain channel into a quasi-time-invariant channel in the DD domain, which can potentially be exploited for striking a compelling trade-off between the performance, computational complexity and signaling overhead.}
%
%

Apart from its potential of exploiting full diversity and Doppler-resilience, OTFS also has some further benefits over conventional modulation techniques.
For example, despite its multicarrier nature, the peak-to-average power ratio (PAPR) of OTFS is much lower than that of both OFDM and generalized frequency division multiplexing (GFDM), which is particularly beneficial for power-limited systems, such as the Internet-of-Things (IoT).
Additionally, the guard intervals are only required between consecutive OTFS frames, rather than between time slots and thus the associated idle time is significantly reduced.
Furthermore, as a benefit of its Doppler-resilience, OTFS is more robust against the carrier frequency offset than OFDM.
These advantages render OTFS eminently suitable for high-mobility and high-carrier scenarios.

\subsection{OTFS System Design Principle}
{According to Fig. \ref{DDDomainMultiplexing}, the data rate of OTFS systems is determined by the amount of data symbols accommodated within a single OTFS frame, with its largest possible value being $MN$.
Since a single OTFS frame occupies a time duration of $NT$ and a bandwidth of $M\Delta f$, the spectral efficiency of OTFS systems is $(1-\eta)R_{\mathrm{c}}\log_2\mathcal{M} $ bit/s/Hz, where $R_{\mathrm{c}}$ is the code rate, $\mathcal{M}$ is the modulation order, and $\eta$ captures the training overhead as well as the guard interval overhead.
The guard interval in the TD has to be longer than the channel's delay spread to avoid interference between OTFS frames.
Furthermore, the code rate $R_{\mathrm{c}}$ and the modulation order $\mathcal{M}$ have to be carefully selected to guarantee communication reliability.
As such, a key technique of improving the spectral efficiency of OTFS systems is that of reducing the training overhead, as shown in Section IV-A.
On the other hand, the OTFS system parameters $M$, $N$, $\Delta f$, and $T$ have to be chosen appropriately for its practical implementations.
In particular, the OTFS frame duration $NT$ has to be smaller than the tolerable application latency.
Furthermore, as the computational complexity and PAPR of OTFS modulation are proportional to $N$ \cite{RavitejaOTFS}, we prefer to choose a smaller $N$ with a longer slot duration $T$.
Meanwhile, the available spectrum is granulated into more subcarriers, i.e. a higher $M$ is associated with a smaller subcarrier spacing $\Delta f = \frac{1}{T}$.}

\section{Potential Applications and Opportunities}
%
OTFS modulation is envisioned to have diversified applications in next-generation wireless networks, as discussed in the following.

\subsection{Vehicular Networks}
Vehicular networks allow various vehicles to wirelessly exchange information with each other or with roadside units (RSUs) to provide a variety of benefits, including cooperative traffic management, road-safety improvements and the support of autonomous driving.
OTFS can play a key role in future vehicular networks owing to its intrinsic advantages for communications over high-mobility channels.
The current standards of vehicular communications, such as IEEE 802.11bd and 5G NR V2X, mainly consider OFDM-based waveforms, where the impact of channel variations is mitigated either by inserting a midamble or by increasing the subcarrier spacing.
By contrast, the potential channel stability of OTFS exhibited in the DD domain enables prompt initial link setup, agile sidelink scheduling, as well as predictive resource scheduling.
Furthermore, the OTFS waveform enjoys a lower PAPR than OFDM, allowing a better communication coverage for vehicular networks.
%

\subsection{Millimeter-wave Communications}
The millimeter-wave frequency band possesses a large amount of under-utilized spectrum and has the potential of offering giga-bit-per-second communication services in future wireless networks.
The Doppler effect becomes more severe upon increasing the carrier frequency even at a low/medium velocity.
Although increasing the subcarrier spacing to mitigate the resultant ICI is feasible, the TD symbol duration will be shorter and inserting a CP for guarding against ISI will introduce a significant overhead.
The excessive phase noise associated with high-frequency oscillators also results in a time-varying composite channel.
OTFS provides a strong immunity to the oscillator phase noise, which is crucial for mmWave communications.

\subsection{Non-Terrestrial Networks}
Non-terrestrial networks (NTN) provide a new telecommunication infrastructure based on airborne or spaceborne vehicles, such as satellites, UAVs or high altitude platforms (HAPs).
They are capable of supporting the terrestrial 5G networks in the provision of global coverage and mobility, as well as ubiquitous connectivity and enhanced network reliability.
Since the airborne and spaceborne vehicles usually move fast, the high Doppler spread experienced by the NTN imposes new challenges on its air interface design.
OTFS modulation has rich potentials in the NTN owing to its prominent capability of handling the Doppler effect.
Additionally, airborne and spaceborne vehicles have limited on-board power supply and computing capability, hence the low PAPR and low-complexity of OTFS are of pivotal importance.
Moreover, the corresponding NTN communication links spanning to the ground terminals usually exhibit spatial channel sparsity in the DD domain, which allows OTFS to strike an attractive performance vs. complexity trade-off.

\subsection{Underwater Acoustic Communications}
Underwater acoustic (UWA) channels are regarded as one of the most challenging wireless channels, due to their high delay spread, limited bandwidth, and rapid time variations.
Single-carrier modulation using decision feedback equalizers (DFE), OFDM and orthogonal signal division multiplexing (OSDM) are the most popular schemes for UWA communications.
However, they all transmit information in the TF domain, where both the ISI and ICI equalization become tedious tasks.
%
%
By contrast, OTFS is Doppler-resilient, hence transmitting information in the DD domain may outperform these TF domain modulation schemes in UWA channels.
Furthermore, UWA channels tend to be sparse in the DD domain, where the equalization might be easier than that in the TF domain.
Moreover, UWA communications are usually considered as wideband systems, since the ratio of acoustic signal bandwidth over the carrier frequency is typically much higher than that in terrestrial communications.
Hence, the potential multipath-scale diversity of the DD domain channel \cite{hlawatsch2011wireless} can be beneficially exploited.

\section{Challenges and Solutions}
As a fledgling waveform, OTFS modulation unveils new opportunities but also has its own challenges.
In this section, we introduce three fundamental research problems of OTFS and their potential solutions.

\subsection{Channel Estimation}
\begin{figure}[!t]
	\centering
	\includegraphics[width=3.5in]{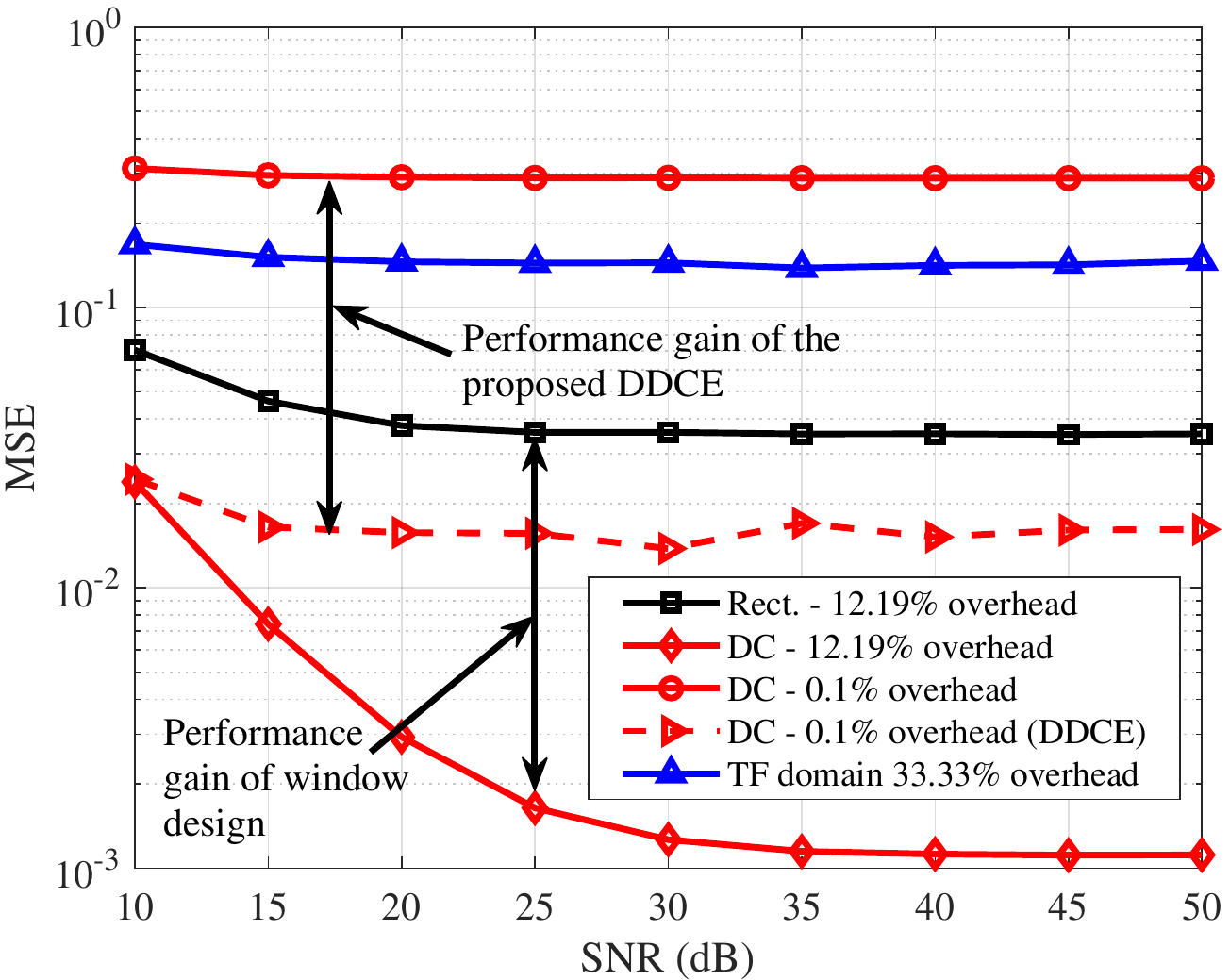}
	\caption{Performance comparison of the DD and TF domain channel estimation with different training overheads. The number of paths in the DD domain is 5, $N = 32$, $M = 32$, $\nu_{\mathrm{max}}$ = $\frac{2}{NT}$, and $\tau_{\mathrm{max}}$ = $\frac{4}{M\Delta f}$.}%
	\label{CEOTFS}%
\end{figure}

The channel envelope fluctuates violently even in a short time period in high-mobility environments.
Accurately estimating the CIR in OTFS systems is a challenging but vital requirement for reliable detection.
Thanks to the DD domain channel sparsity and quasi-stationarity, channel acquisition in the DD domain is more convenient than that in the TF domain, even for a lower training overhead, cf. Fig. \ref{CEOTFS}.
However, the DD domain channel may not always be sparse\cite{WeiOTFS}, particularly in the presence of fractional Doppler\cite{RavitejaOTFS}.
For example, when the exact Doppler frequency straddles a pair of finite-resolution bins in Fig. \ref{DDDomainMultiplexing}, rather than falling exactly into the $1$-th bin, the DD domain channel is spread across all
the Doppler indices.
Due to this channel spreading, a much larger guard space is needed around the pilot symbols to avoid the interference caused by unknown data symbols for channel estimation, which imposes a significant training overhead.

A promising solution to address this issue is enhancing the channel sparsity via designing a bespoke TF domain window.
In particular, we proposed to apply a Dolph-Chebyshev (DC) window at the OTFS transmitter or receiver \cite{WeiOTFS} to suppress the channel spreading.
Due to the enhanced channel sparsity, the DC windowing achieves a much improved channel estimation accuracy over the conventional rectangular window \cite{RavitejaOTFSCE}, cf. Fig. \ref{CEOTFS}.
{To further reduce the training overhead, we propose to use decision-directed channel estimation (DDCE), where the reliably detected data symbols can be used for refining channel estimates based on the classic decision-directed principle, rather than purely relying on the known pilot symbols.
	The refined channel estimates can again be used for OTFS symbol detection, which can in turn improve the channel estimation accuracy.
	Fig. \ref{CEOTFS} demonstrates that the proposed DDCE scheme without guard space consumes only $0.1\%$ training overhead while can achieve an order of magnitude reduction in channel estimation error than that of its conventional counterpart.}

%

\subsection{Efficient DD Domain Data Detection}
The output signal in the DD domain can be regarded as a 2D circular convolution of the input data symbols and the effective aggregate channel, cf. Fig \ref{DDDomainMultiplexing}, which results in a rather specific interference pattern, where a pair of symbols far from each other in the DD domain may interfere with each other.
Mitigating this peculiar interference requires a bespoke receiver.
Adopting the optimal maximum a posteriori (MAP) detector would indeed perfectly mitigate the interference between symbols, but at an excessive complexity, precluding its deployment in practical systems. 
Hence, most OTFS detectors focused on the complexity reduction, based on the classic message passing algorithm (MPA) and its variants. 
The main problem of MPA-based detection is its poor convergence behavior in the face of short cycles, which may lead to performance degradation.

\begin{figure}[!t]
	\centering
	\includegraphics[width=3.5in]{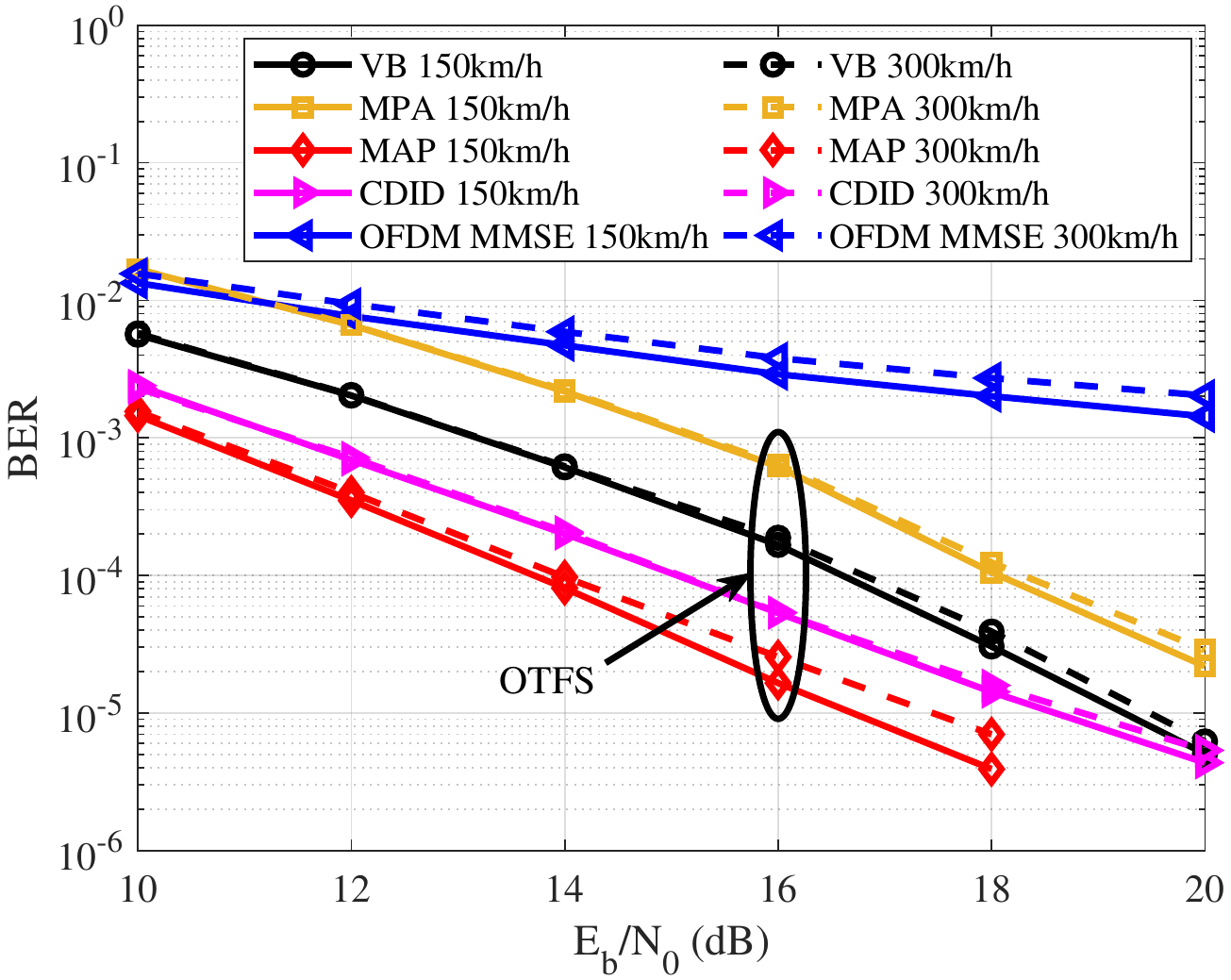}
	\caption{Performance comparison of OTFS with different detectors, moving speeds, and waveforms. The number of paths in the DD domain is 4, $N = 8$, $M = 16$, $\nu_{\mathrm{max}}$ = $\frac{3}{NT}$ for the velocity of $150$ km/h, $\nu_{\mathrm{max}}$ = $\frac{6}{NT}$ for the velocity of $300$ km/h, and $\tau_{\mathrm{max}}$ = $\frac{3}{M\Delta f}$.}%
	\label{Detection}%
\end{figure}

A potent solution is to adopt the variational framework of\cite{WeijieOTFS}, which can adaptively construct the distributions of OTFS symbols according to their interference patterns. 
By appropriately constructing the distributions of OTFS symbols for variational purposes, we can design rapidly converging OTFS detection.
Owing to its better convergence, the variational Bayes (VB) OTFS detector can achieve a modest performance gain over the MPA detector, cf. Fig. \ref{Detection}.
The performance of MAP detection is also provided as the baseline, which has the best performance, albeit at the cost of an excessive complexity. 
{Another potential solution is the so-called cross-domain iterative detection (CDID), where a conventional linear minimum mean squared error (L-MMSE) estimator is adopted for equalization in the time domain and low-complexity symbol-by-symbol detection is utilized in the DD domain, while the extrinsic information is iteratively exchanged between the time domain and DD domain via the corresponding unitary transformation.
The proposed CDID detector is also capable of exploiting the time domain channel sparsity and the DD domain symbol constellation constraints, which can achieve a close-to-optimal performance at a much reduced computational complexity, cf. Fig. \ref{Detection}.}
For all these detectors, the OTFS performance remains similar upon increasing the velocity from 150 km/h to 300 km/h.
In contrast, the detection performance of the MMSE OFDM detector remains poor due to the excessive ICI.

%
%
%

%
%

\subsection{Coded OTFS System}

\begin{figure}[!t]
	\centering
	\includegraphics[width=3.5in]{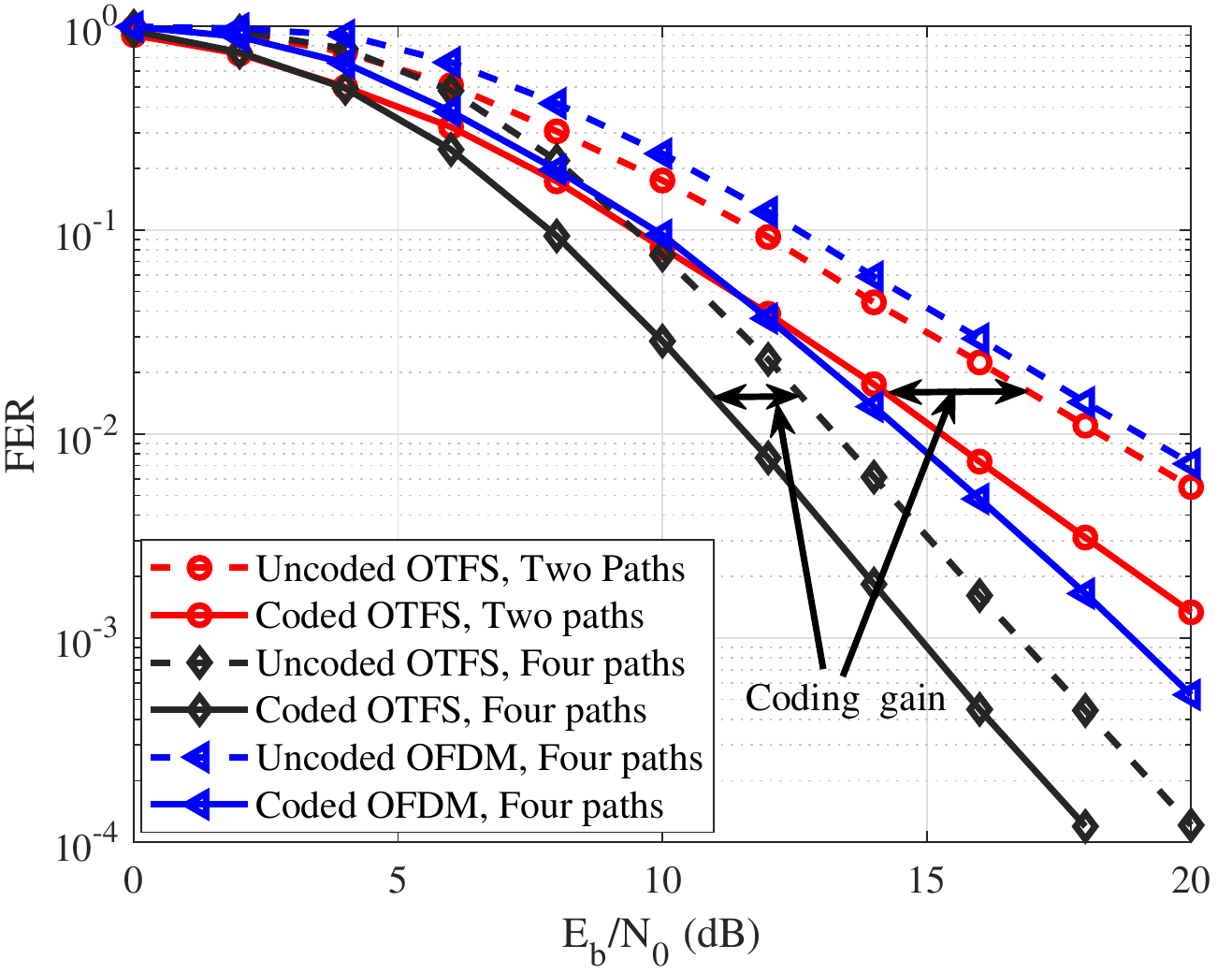}
	\caption{Performance comparison of coded/uncoded OTFS and OFDM modulations, $N = 8$, $M = 16$, $\nu_{\mathrm{max}}$ = $\frac{3}{NT}$, and $\tau_{\mathrm{max}}$ = $\frac{5}{M\Delta f}$.}%
	\label{Code_OTFS}%
\end{figure}

{
%
%
While OTFS has the potential of attaining the maximum achievable diversity gain, the channel codes have to be carefully designed for OTFS modulation. 
%
%
%
Moreover, perfect detection at near-capacity signal-to-noise ratios (SNRs) may not be attained for practical OTFS systems due to the associated poor channel conditions. 
Hence, the channel decoder has to cope with the OTFS detector's residual errors, which would require iterative OTFS receivers.
However, how to design such a receiver and how to choose the coding parameters for near-capacity joint detection and decoding remains an interesting open issue.}
Recent research has unveiled a fundamental trade-off between the diversity gain and coding gain, which may shed light on code design of OTFS systems \cite{ShuangyangOTFS}.
In particular, the diversity gain of OTFS systems improves with the number of independent resolvable channel paths in the DD domain, while the coding gain declines, cf. Fig. \ref{Code_OTFS}.
{This is not unexpected as the transmitted signal energy is distributed to multiple paths, as discussed in \cite{ShuangyangOTFS}.}
Moreover, both the coded and uncoded OTFS systems outperform the corresponding OFDM systems, which clearly shows the advantage of OTFS modulation. 

%
%
%
%
%

%

\section{Outlook for OTFS}
OTFS has great potential in providing reliable communications for high-mobility applications and its open facets are expected to stimulate new research, as discussed below.


\subsection{FDD or TDD?}
As demonstrated by Cohere \cite{Hadani2018otfs}, the DD domain channels are capable of reaching a coherence time of 100 ms and a coherence bandwidth of 100 MHz.
Hence, accurate channel reciprocity holds in the DD domain and thus time-division duplexing (TDD) is a good option for OTFS.
Besides, due to the significantly increased coherence bandwidth of the DD domain channel, we can potentially infer the downlink channel from the uplink channel even for frequency-division duplexing (FDD) OTFS systems, which significantly reduces the channel feedback overhead.
However, the key is to find an accurate deterministic or statistical mapping between the uplink and downlink channels.
Furthermore, it is also important to conduct a systematic comparative study of TDD and FDD-aided OTFS systems.

\subsection{Scalable Multiple Access Schemes for OTFS}
How to support a multiplicity of users in high-mobility environments is a very challenging issue.
OTFS offers the opportunity to accommodate multiple users in the DD domain, where employing carefully designed user scheduling and guard spaces has the potential of avoiding multi-user interference.
However, how to scale the systems for accommodating a large number of users without a significant overhead is an interesting open research problem.
The coexistence of promising multiple access schemes and OTFS, such as non-orthogonal multiple access, spatial-division multiple access and interleave-division multiple access, is worth further exploring.

\subsection{Joint Sensing and Communications using OTFS}
Since the DD domain channel directly exploits the physics of propagation, relying on the distance, speed and scattering intensity, OTFS is eminently suitable for integrating sensing and communications solutions in a single platform.
Efficient sensing algorithms to exploit the OTFS signal structure are still unknown.
Finding the optimal trade-off between the sensing and communication performances remains an interesting open question.
Moreover, as location and velocity can serve as beneficial side information for improving communication performance, sensing-based communications relying on OTFS is an exciting open topic to investigate.

\subsection{MIMO-OTFS}
Applying OTFS in multi-antenna systems provides additional hitherto unexploited spatial DoF for multiplexing.
In contrast to TF domain channels, which may fluctuate dynamically for different antennas in different time slots and subcarriers, the DD domain channels tend to remain quasi-stationary both in the time and antenna domains, which may result in an efficient channel estimation and multi-input multiple-input multiple-output (MIMO) detection.
How to design sophisticated beamforming/precoding to fully exploit all the available spatial DoFs and how to perform low-complexity detection for MIMO-OTFS constitute intriguing problems.
Moreover, the analytical framework of MIMO-OTFS system performance versus the number of antennas is also unexplored in the open literature.

\subsection{Index Modulation for OTFS}
{OTFS modulation maps the classic modulated symbols to the DD domain and spreads them across the whole TF domain grid for transmission.
Therefore, IM can be carried out in the TF domain for improving the spectral efficiency, while slightly sacrificing the transmission diversity order.
In particular, the additional information bits can be mapped to the ON/OFF states of the TF grid indices.
Additionally, we can also extend this to the spatial dimension, such as antenna indices or propagation paths, for incorporating IM into OTFS systems, where the information bits are transmitted by active antenna or path indices.
The intriguing error rate performance vs.  complexity and spectral efficiency trade-offs of IM-OTFS systems remain to be investigated.
Furthermore, the design of efficient detectors for jointly demodulating the additional information bits and conventional constellation symbols of IM-OTFS systems is also a critical open research challenge.}

\section{Conclusions}
In a nutshell, OTFS constitutes a promising next-generation candidate.
We commenced with an overview of the fundamental concept of OTFS, including the main features of the DD domain channel, the DD domain multiplexing and OTFS transceiver architecture.
The critical challenges of OTFS, such as channel estimation, efficient data detection and coding/decoding problems were highlighted and pertinent preliminary results were provided.
The potential applications of OTFS and several promising research directions were introduced.
It is hoped that this article will help inspire future research in this exciting new area and pave the way for designing next-generation networks.

\section{Acknowledgement}
The authors would like to thank the support from Telstra Corporation Ltd., particularly Dr. Paul G. Fitzpatrick, Dr. Taka Sakurai, and Mr. Paul Sporton for valuable discussions during this work.

\bibliographystyle{IEEEtran}
\bibliography{OTFS}

\end{document}